\documentstyle[12pt]{article}
\begin{document}
\title{Mapping the proton drip line
from \\$Z=31$ to $Z=49$}
\author{G.A. Lalazissis$^{1,2}$, D. Vretenar$^{1,3}$, and P. Ring$^{1}$
\vspace{0.5 cm}\\
$^{1}$ Physik-Department der Technischen Universit\"at M\"unchen,\\
D-85748 Garching, Germany\\
$^{3}$ Physics Department, Aristotle University of Thessaloniki,\\
Thessaloniki GR-54006, Greece\\
$^{3}$ Physics Department, Faculty of Science, 
\\University of Zagreb, 10 000 Zagreb, Croatia}
\maketitle
\bigskip
\bigskip
\begin{abstract}
The structure of proton drip line nuclei in the 
$60 < A < 100$ mass range is studied with the 
Relativistic Hartree Bogoliubov (RHB) model. 
For the elements which determine the astrophysical 
rapid proton capture process path, the RHB model 
predicts the location of the proton drip-line,
the ground-state quadrupole deformations and
one-proton separation energies at and beyond the drip-line.
The results of the present theoretical investigation
are compared with available experimental data.
For possible odd-Z ground state proton
emitters, the calculated deformed single-particle
orbitals occupied by the odd valence proton 
and the corresponding spectroscopic factors are 
compared with predictions of the macroscopic-microscopic 
mass model.
%#####################################################
\end{abstract}
%#####################################################
\bigskip \bigskip

%#####################################################
\vspace{1 cm} {PACS:} {21.60.Jz, 21.10.Dr, 21.10.Jx, 26.50.+x,
27.50.+e}\newline
\vspace{1 cm}\newline
\newpage
\baselineskip = 24pt

%=========================================================================
%  Section 1
%
\section{Introduction and outline of the relativistic Hartree-Bogoliubov
model}
%=========================================================================
The structure of nuclei at the proton drip line in the mass region 
$60 < A < 100$ is important for the process of nucleosynthesis 
during explosive hydrogen burning. The exact location of the
proton drip line determines a possible path of rapid proton capture
process. The path of the rp-process lies between the line of 
$\beta$-stability and the drip line, and it is a very complicated
function of the physical conditions, temperature and density,
governing the explosion~\cite{CW.92}. The input for rp-process 
nuclear reaction network calculations includes the nuclear 
masses, or proton separation energies of the neutron deficient 
isotopes, the proton capture rates, their inverse photodisintegration
rates, the $\beta$-decay and electron capture rates. In a recent
extensive analysis~\cite{Sch.98}, the influence of nuclear structure
on the rp-process between Ge and Sn at extreme temperature
and density conditions, has been studied with a number 
of theoretical models for the nuclear masses, deformations, 
reaction rates and $\beta$-decay rates.

In addition to its importance for astrophysical processes, 
the information about the exact location of the drip line, 
as well as the proton separation 
energies beyond the drip line, are essential for studies 
of ground state proton radioactivity~\cite{WD.97}.
No examples of ground state proton emitters below $Z=50$
have been reported so far, and therefore theoretical 
studies might provide important information for future
experiments in this region.

The proton drip line has been fully mapped up to Z=21, and
possibly for odd-Z nuclei up to In \cite{WD.97}. In Ref.~\cite{Orm.97}
the proton drip line has been mapped up to $A=70$ by 
calculating Coulomb energy differences between mirror nuclei within 
the framework of the nuclear shell model. The structure of 
proton drip line nuclei around the doubly magic $^{48}$Ni has 
been studied with various self-consistent mean field models
in Ref.~\cite{Naz.96}, and the proton capture reaction
rates in the mass range $23 \leq A \leq 43$ have been 
calculated with the nuclear shell model in Ref.~\cite{HGW.95}.
In a recent series of articles~\cite{VLR.98,VLR.99,LVR.99,LVR.99a},
we have applied the Relativistic Hartree Bogoliubov (RHB) 
model~\cite{Rin.96,PVL.97,LVP.98} in the study of ground state 
properties of proton rich nuclei. In general, the calculated
properties have been found in excellent agreement with 
available experimental data, and with the predictions
of the macroscopic-microscopic mass model. In the present 
study we apply the RHB model in the description of 
the structure of nuclei along the proton drip line from 
$Z=31$ to $Z=49$. We will determine 
the location of the proton drip-line, the ground-state quadrupole
deformations and one-proton separation energies at and beyond the drip-line,
the deformed single-particle orbitals occupied by the odd valence proton
in odd-Z nuclei, and the corresponding spectroscopic factors.

A detailed discussion of the relativistic Hartree-Bogoliubov
theory can be found, for instance, in Ref. \cite{LVR.99}.
For completeness we include a description of the essential
features of the RHB model. In Section II we present and discuss
the results of the analysis the proton drip line from 
$Z=31$ to $Z=49$. Section III contains the summary.

In the framework of the relativistic mean field theory 
nucleons are described as point particles that
move independently in mean fields
which originate from the nucleon-nucleon interaction.
The theory is fully Lorentz invariant.
Conditions of causality and Lorentz invariance impose that the
interaction is mediated by the
exchange of point-like effective mesons, which couple to the nucleons
at local vertices. The single-nucleon dynamics is described by the
Dirac equation
\begin{equation}
\label{statDirac}
\left\{-i\mbox{\boldmath $\alpha$}
\cdot\mbox{\boldmath $\nabla$}
+\beta(m+g_\sigma \sigma)
+g_\omega \omega^0+g_\rho\tau_3\rho^0_3
+e\frac{(1-\tau_3)}{2} A^0\right\}\psi_i=
\varepsilon_i\psi_i.
\end{equation}
$\sigma$, $\omega$, and
$\rho$ are the meson fields, and $A$ denotes the electromagnetic potential.
$g_\sigma$ $g_\omega$, and $g_\rho$ are the corresponding coupling
constants for the mesons to the nucleon.
The lowest order of the quantum field theory is the {\it
mean-field} approximation: the meson field operators are
replaced by their expectation values. The sources
of the meson fields are defined by the nucleon densities
and currents.  The ground state of a nucleus is described
by the stationary self-consistent solution of the coupled
system of the Dirac~(\ref{statDirac})and Klein-Gordon equations:
\begin{eqnarray}
\left[ -\Delta +m_{\sigma }^{2}\right] \,\sigma ({\bf r}) &=&-g_{\sigma
}\,\rho _{s}({\bf r})-g_{2}\,\sigma ^{2}({\bf r})-g_{3}\,\sigma ^{3}({\bf r})
\label{messig} \\
\left[ -\Delta +m_{\omega }^{2}\right] \,\omega ^{0}({\bf r}) &=&g_{\omega
}\,\rho _{v}({\bf r})  \label{mesome} \\
\left[ -\Delta +m_{\rho }^{2}\right] \,\rho ^{0}({\bf r}) &=&g_{\rho }\,\rho
_{3}({\bf r})  \label{mesrho} \\
-\Delta \,A^{0}({\bf r}) &=&e\,\rho _{p}({\bf r}),  \label{photon}
\end{eqnarray}
for the sigma meson, omega meson, rho meson and photon field, respectively.
Due to charge conservation, only the 3rd-component of the isovector rho
meson contributes. The source terms in equations (\ref{messig}) to (\ref
{photon}) are sums of bilinear products of baryon amplitudes, and they 
are calculated in the {\it no-sea} approximation, i.e. the Dirac sea
of negative energy states does not contribute to the nucleon densities
and currents. Due to time reversal invariance,
there are no currents in the static solution for an even-even
system, and therefore the spatial
vector components \mbox{\boldmath $\omega,~\rho_3$} and
${\bf  A}$ of the vector meson fields vanish.
The quartic potential 
\begin{equation}
U(\sigma )~=~\frac{1}{2}m_{\sigma }^{2}\sigma ^{2}+\frac{1}{3}g_{2}\sigma
^{3}+\frac{1}{4}g_{3}\sigma ^{4}  \label{usigma}
\end{equation}
introduces an effective density dependence. The non-linear
self-interaction of the $\sigma$ field is essential for 
a quantitative description of properties of finite nuclei.

In addition to the self-consistent mean-field
potential, pairing correlations have to be included in order to
describe ground-state properties of open-shell nuclei.
For nuclei close to the $\beta$-stability
line, pairing can been included in the relativistic
mean-field model in the form of a simple BCS
approximation. For more exotic nuclei further
away from the stability line, however, the BCS model presents 
only a poor approximation. In particular, in order to 
correctly reproduce density distributions in nuclei close to  the 
drip lines , mean-field and pairing correlations have to be 
described in a unified framework: the Hartree-Fock-Bogoliubov
model or the relativistic Hartree-Bogoliubov (RHB) model.
In the unified framework 
the ground state of a nucleus $\vert \Phi >$ is represented
by the product of independent single-quasiparticle states.
These states are eigenvectors of the
generalized single-nucleon Hamiltonian which
contains two average potentials: the self-consistent mean-field
$\hat\Gamma$ which encloses all the long range particle-hole ({\it ph})
correlations, and a pairing field $\hat\Delta$ which sums
up the particle-particle ({\it pp}) correlations.
In the Hartree approximation for
the self-consistent mean field, the relativistic
Hartree-Bogoliubov equations read
\bigskip
\begin{eqnarray}
\label{equ.2.2}
\left( \matrix{ \hat h_D -m- \lambda & \hat\Delta \cr
		-\hat\Delta^* & -\hat h_D + m +\lambda} \right)
		\left( \matrix{ U_k({\bf r}) \cr V_k({\bf r}) } \right) =
		E_k\left( \matrix{ U_k({\bf r}) \cr V_k({\bf r}) } \right).
\end{eqnarray}
where $\hat h_D$ is the single-nucleon Dirac
Hamiltonian (\ref{statDirac}), and $m$ is the nucleon mass.
The chemical potential $\lambda$  has to be determined by
the particle number subsidiary condition in order that the
expectation value of the particle number operator
in the ground state equals the number of nucleons. The column
vectors denote the quasi-particle spinors and $E_k$
are the quasi-particle energies.
The pairing field $\hat\Delta $ in (\ref{equ.2.2}) is defined
\begin{equation}
\label{equ.2.5}
\Delta_{ab} ({\bf r}, {\bf r}') = {1\over 2}\sum\limits_{c,d}
V_{abcd}({\bf r},{\bf r}') \sum_{E_k>0} U_{ck}^*({\bf r})V_{dk}({\bf r}'),
\end{equation}
where $a,b,c,d$ denote quantum numbers
that specify the Dirac indices of the spinors,
$V_{abcd}({\bf r},{\bf r}')$ are matrix elements of a
general two-body pairing interaction.
The RHB equations are solved self-consistently, with
potentials determined in the mean-field approximation from
solutions of Klein-Gordon equations for the meson fields.
The current version of the model~\cite{LVR.99}
describes axially symmetric deformed shapes.
The Dirac-Hartree-Bogoliubov equations and the equations for the
meson fields are solved by expanding the nucleon spinors
$U_k({\bf r})$ and $V_k({\bf r})$,
and the meson fields in terms of the eigenfunctions of a
deformed axially symmetric oscillator potential.
A simple blocking procedure is used in the calculation of
odd-proton and/or odd-neutron systems. The blocking calculations
are performed without breaking the time-reversal symmetry.

The input parameters of the RHB model are the coupling constants and the
masses for the effective mean-field Lagrangian, and the effective
interaction in the pairing channel. In most applications we have
used  the NL3 effective interaction \cite{LKR.97} for the RMF
Lagrangian. Properties calculated with NL3 indicate that this is probably
the best effective interaction so far, both for nuclei at and away from the
line of $\beta $-stability. For the pairing field we employ the
pairing part of the Gogny interaction
\begin{equation}
V^{pp}(1,2)~=~\sum_{i=1,2}
e^{-(( {\bf r}_1- {\bf r}_2)
/ {\mu_i} )^2}\,
(W_i~+~B_i P^\sigma
-H_i P^\tau -
M_i P^\sigma P^\tau),
\end{equation}
with the set D1S \cite{BGG.84} for the parameters
$\mu_i$, $W_i$, $B_i$, $H_i$ and $M_i$ $(i=1,2)$.
This force has been very carefully adjusted to the pairing
properties of finite nuclei all over the periodic table.
In particular, the basic advantage of the Gogny force
is the finite range, which automatically guarantees a proper
cut-off in momentum space.
%=========================================================================
%  Section 2
%
\section{The proton drip line from Z=31 to Z=49}
%=========================================================================

In this section, we use the relativistic Hartree-Bogoliubov model
to map the proton drip line from $Z=31$ to $Z=49$. As in our 
previous studies of proton-rich nuclei~\cite{VLR.98,VLR.99,LVR.99,LVR.99a},
the NL3 effective interaction is used in the mean-field Lagrangian,
and pairing correlations are described by the 
pairing part of the finite range Gogny interaction D1S. We analyze the
structure of proton drip line nuclei in the 
$60 < A < 100$ mass range: 
the location of the proton drip-line, the ground-state quadrupole
deformations and one-proton separation energies at and beyond the drip-line,
the deformed single-particle orbitals occupied by the odd valence proton
in odd-Z nuclei, and the corresponding spectroscopic factors.

In Fig. \ref{figA} we display the section of the chart of the nuclides
along the proton drip line in the region  $31 \leq Z \leq 49$. The 
calculation predicts the last bound isotopes for each element.
Nuclei to the left are proton unstable. For odd-Z nuclei the 
proton drip line can be compared with available experimental 
data. For $Z=31$ and $Z=33$ the calculated drip line nuclei 
$^{61}$Ga and $^{65}$As, respectively, are in agreement with 
experimental data reported in Refs.~\cite{Moh.91,Win.93}. 
These two nuclei are on the on the rp-process path proposed
by Champagne and Wiescher~\cite{CW.92}. In Ref.~\cite{Bla.95}
evidence was reported for the existence of $^{60}$Ga, but 
of course the observation of an isotope does not necessarily 
imply that the nucleus is proton bound, but rather that 
its half-life is longer than the flight time through 
the fragment analyzer. 

For $Z=35$ the RHB calculation predicts that the last proton
bound isotope is $^{70}$Br. The isotope $^{69}$Br is 
calculated to be proton unbound in most mass models (see also
Fig. \ref{figB}). Experimental evidence for $^{69}$Br was
reported in Ref.~\cite{Moh.91}, but no evidence for this
isotope was found in the experiment of Ref.~\cite{Bla.95}, 
and it was deduced that $^{69}$Br is proton unbound with 
a half-life shorter than about 100 ns. 

For $Z=37$ the experiment of Ref.~\cite{Moh.91} confirms that
$^{74}$Rb is the last proton bound nucleus, in agreement with
the result of the present calculation. For $Z=39,41,43$ the
lightest isotopes observed in the experiment of 
Ref.~\cite{Yen.92} are $^{78}$Y, $^{82}$Nb and $^{86}$Tc,   
respectively. While for Nb and Tc these results correspond
to the drip line as calculated in the present work, for Y 
the RHB model predicts that the last proton bound nucleus
is $^{77}$Y. This isotope would then be the heaviest 
$T_z = - {1\over 2}$ nucleus, in contrast to the suggestion
of Ref.~\cite{Moh.91} that $^{69}$Br is most likely the
highest observable odd-Z $T_z = - {1\over 2}$ nucleus.

The calculated odd-Z drip line nuclei $^{90}$Rh and $^{94}$Ag
were observed in the experiment reported in Ref.~\cite{Hen.94},
and experimental evidence for $^{98}$In was reported in 
Ref.~\cite{Ryk.95}.

For even-Z nuclei in this region, it was not possible to 
compare the calculated proton drip line with experimental
data. While for the odd-Z elements most of the last proton 
bound nuclei lie on the $N=Z$ line, with just few $T_z = - {1\over 2}$
nuclei, the proton drip line for even-Z elements is calculated
to be at $T_z = - 3$, or even at $T_z = - {7\over 2}$.
The only exception is the drip line nucleus $^{84}$Ru 
with $T_z = - 2$. Nuclei with such extreme values of $T_z$
are virtually impossible to produce in experiments, and since they
lie so far away from the rp-process path, the even-Z proton drip line
nuclei in this mass region play no role in the process of 
nucleosynthesis during explosive hydrogen burning. We have, however,
compared our calculated proton drip line with the predictions
of other, well known and frequently used 
mass models~\cite{Hil.76,JM.88,MN.95,MNK.97}. The comparison
is illustrated in Fig. \ref{figB}, where the 
mass number of the first proton unbound nucleus along each
isotopic chain is plotted as a function of the atomic number.
We notice that, with the exception of the classical
mass formula by Hilf et al.~\cite{Hil.76}, all models agree
on the location of the proton drip line for odd-Z nuclei 
(see also Fig. 9 in Ref.~\cite{Sch.98}). The only significant
difference is $^{77}$Y which, in contrast to the mass models of 
Refs.~\cite{JM.88,MN.95,MNK.97}, is predicted to be
the heaviest proton bound $T_z = - {1\over 2}$ nucleus by
the present RHB/NL3 calculation.
This difference could be important because, if $^{77}$Y
were bound and therefore located on the rp-process path, $^{76}$Sr
would not be a waiting point nucleus. For even Z-nuclei, the 
theoretical models differ in their predictions for the location of 
the proton drip line. As it is shown in Fig. \ref{figB}, the 
differences are especially pronounced for Sr, Zr, Ru and Pd,
and they reflect the different treatment of pairing correlations
and deformation effects. In fact, the combined effect of 
pairing correlations and nuclear deformation is responsible
for the large difference between the $T_z$ values 
for even-Z and odd-Z nuclei at the proton drip line. 
The calculated  ground-state quadrupole deformations of the 
last proton bound nuclei are shown in Fig. \ref{figC}. 
For $Z \leq 33$ the drip line nuclei are moderately deformed,
between $34 \leq Z \leq 41$ the odd-Z drip line nuclei are 
highly deformed, and for $Z > 41$ (protons in the $g~9/2$ orbital)
the drip line enters a region of spherical nuclei. 
Between $34 \leq Z \leq 41$ one notices that the odd-Z nuclei 
are much more deformed than their even-Z neighbors
on the drip line. The reason is that pairing correlations are 
strongly reduced in odd-Z nuclei, and as a result the nucleus
is driven toward larger deformations. Much stronger pairing 
in even-Z nuclei results in almost spherical shapes, which
in turn shift the drip line to extremely low values 
$T_z \approx -3$. One could also say that the strong reduction
of pairing in odd-Z nuclei causes the drip line to lie at 
$T_z = - {1\over 2}$ or $T_z = 0$. We have verified that the 
blocking of odd proton orbitals is essential for the 
correct description of the drip line in odd-Z nuclei. Without
blocking, the calculated drip line in Fig. \ref{figA} is shifted to
the left, to the position of the drip line of even-Z nuclei.

An important issue in future experimental studies of proton
rich nuclei in this mass region is the possible observation
of ground state proton emission. 
The structure and decays modes of nuclei beyond the proton drip-line
represent one of the most active areas of experimental and theoretical
studies of exotic nuclei with extreme isospin values. In the last
few years many new data on ground-state and isomeric proton
radioactivity have been reported in the region 
$51 \leq Z \leq 83$. In particular, measured half-lives and 
transition energies, as well as calculated transition rates, have
shown that the ground state proton emitters are spherical 
in the regions 51$\leq $Z$\leq $55 
and 69$\leq $Z$\leq $ 83~\cite{WD.97}, and strongly deformed
in the region of light rare-earth nuclei 
57$\leq $Z$\leq $67~\cite{Dav.98,Ryk.99}.
In Refs.~\cite{VLR.99,LVR.99,LVR.99a} we have applied 
the RHB (NL3+D1S) model in the description of 
ground-state properties of proton-rich odd-Z
nuclei in the region $53 \leq Z \leq 71$. We have 
calculated proton separation energies, ground-state quadrupole
deformations, single-particle orbitals occupied by the odd
valence proton, and the corresponding spectroscopic factors.
The results were found to be in excellent agreement with
experimental data, and we were also able to predict several
new ground state proton emitters. And while the
relatively high potential energy barrier enables the
observation of ground state proton emission from
medium-heavy and heavy nuclei, no examples of ground state proton
radioactivity have been discovered so far below $Z=50$.
The reason is, of course, the low Coulomb barrier. Nuclei 
beyond the proton drip line in this region 
exist only as short lived resonances. 
For a typical rare-earth nucleus the window of proton energies,
i.e. the $Q_{p}$ values for which ground state proton emission
can be directly observed is about 0.8 -- 1.7 MeV \cite{ASN.97}.
For lower $Q_{p}$ values the total half-life will be completely 
dominated by $\beta ^{+}$ decay; higher transition energies
result in extremely short proton-emission half-lives which 
cannot be observed directly. In an early study of the phenomena
of proton and two-proton radioactivity~\cite{Gol.60}, Goldansky
has used an approximate formula to calculate the half-life
of a ground state proton emitter
\begin{equation}
{\rm log}~T_{1\over 2} (sec) \approx 0.43 Z^{2\over 3} f(x) - 22,
\end{equation}
where x is the ratio of the proton transition energy to the 
height of the Coulomb barrier. For $x << 1$ 
\begin{equation}
f(x) \approx 0.6 ( {1\over 2} \pi x^{-{1\over 2}} - 2).
\end{equation} 
For the interval $T_{1\over 2} = 10 - 10^{-4}$ sec, 
the corresponding energy range of the emitted protons 
is calculated: $0.2 - 0.3$ MeV (for Z=30) and $0.35 - 0.5$
MeV (for Z=40).

In Figs. \ref{figD} and \ref{figE} the one-proton separation energies 
\begin{equation}
S_{p}(Z,N)=B(Z,N)-B(Z-1,N),  \label{sep}
\end{equation}
are displayed for the odd-Z nuclei $31\leq Z\leq 49$, as function of the
number of neutrons. In both figures the energy window extends beyond 
the proton drip line, in order to include those
nuclei for which a direct observation of ground-state proton emission 
is in principle possible on the basis of calculated separation energies. 
For the best candidates, the ground-state properties calculated 
with the RHB (NL3+D1S) model are displayed in Table \ref{TabA}.

The table includes the one-proton separation energies $S_{p}$,
the quadrupole deformations $\beta _{2}$, the deformed 
single-particle orbitals occupied by the odd valence proton, 
and the corresponding theoretical spectroscopic factor. The
spectroscopic factor $u_{\Omega }^{2}$, which results from 
the self-consistent treatment of pairing in the RHB model, 
is defined as the probability that the deformed odd-proton 
orbital is empty in the daughter nucleus with even
proton number. In addition to the isotopes with one-proton 
separation energies in the energy range $\approx$ -0.1 to -0.5
MeV, we have also included a few nuclei with higher transition
energies ($^{72}$Rb, $^{88}$Rh and $^{96}$In). The reason 
is that, for these odd-odd nuclei, we cannot expect the RHB
mean-field model to predict the one-proton separation energies
with high accuracy. The model does not include any residual 
interaction between the odd proton and the odd neutron. Such
an additional interaction, which could be represented for 
instance by the surface delta-force, will increase the binding 
energy. As a result, the proton transition energy would have
a lower value, maybe in the energy range of observable ground
state proton emission. 

In Table \ref{TabA} we also compare the results of the present
calculations with the predictions of the finite-range droplet 
(FRDM) mass model: the projection of the odd-proton angular
momentum on the symmetry axis and the parity of the odd-proton state 
$\Omega_{p}^{\pi }$ \cite{MNK.97}, the one-proton separation 
energy \cite{MNK.97}, and the ground-state quadrupole 
deformation \cite{MN.95}. In general, the two models predict
very similar quadrupole deformations at the drip line. Nuclei
with $Z \leq 41$ display pronounced quadrupole deformations, and  
spherical nuclei at the drip lines are found for $43 \leq Z$.
The only notable difference is $^{73}$Rb. Both models
predict the same one-proton separation energy, but the nucleus is 
calculated to be oblate in the RHB model ($\beta _{2} = -0.34$),
while a prolate shape is obtained in FRDM ($\beta _{2} = 0.37$).
Correspondingly, the two models differ in the prediction of the 
odd-proton orbital occupied in $^{73}$Rb. In fact, 
the predictions for the deformed 
orbitals occupied by the odd proton differ in many cases. 
For the two As isotopes beyond the drip line, the RHB calculation
predicts the $1/2^{-}[310]$ proton orbital, while a 
$3/2^{-}$ orbital is occupied in the FRDM. In the RHB calculation
for $^{85}$Tc and $^{89}$Rh the odd proton is found in the 
$9/2^{+}[404]$ Nilsson orbital, while a $3/2^{+}$ and a $5/2^{+}$
states are, respectively, predicted by the FRDM. The calculated
proton energies beyond the drip line are, on the average, higher
in the FRDM. With the exception of $^{69}$Br, which is calculated
to be proton bound in the FRDM, this model predicts more negative
separation energies in the region of deformed nuclei $Z \leq 41$.
These divergences might be related to the different
descriptions of the effective spin-orbit single-nucleon potential. 

The self-consistent ground-state quadrupole deformations for odd-Z
nuclei $31 \leq Z \leq 47$, at and beyond the proton drip-line, 
are shown in Fig. \ref{figF}. The Ga and As isotopes are moderately
deformed, the Br nuclei are oblate, and a transition
from prolate to oblate shapes is observed at the drip line in Rb. 
The Y isotopes are strongly prolate deformed, shape transitions 
are predicted in Nb, and the isotopes of Tc, Rh and Ag are 
essentially spherical at the drip line.

\bigskip
%=========================================================================
%  Section 3
%
\section{Summary}
%=========================================================================

In this study we have analyzed the structure of the proton drip
line in the mass region $60 < A < 100$. The theoretical model
that we have used is based on the  relativistic Hartree-Bogoliubov
theory. This framework provides a unified and
self-consistent description of mean-field and pairing correlations,
which is especially important for applications to 
exotic nuclei far from the valley of $\beta$-stability. In addition,
the relativistic model includes the important isospin dependence
of the spin-orbit term of the effective single-nucleon
potential. The NL3 effective interaction 
has been used for the mean-field Lagrangian, and pairing correlations
have been described by the pairing part of the finite range Gogny 
interaction D1S. This particular combination of effective forces 
in the $ph$ and $pp$ channels has been used in most of our applications
of the RHB theory. 

The proton drip line nuclei in the mass range $60 < A < 100$ 
determine the astrophysical rapid proton capture process path.
Their properties are, therefore, crucial for a correct
description of the process of nucleosynthesis during 
explosive hydrogen burning. In the present analysis we
have calculated 
the location of the proton drip-line, the ground-state quadrupole
deformations and one-proton separation energies at and beyond the drip-line,
the deformed single-particle orbitals occupied by the odd valence proton
in odd-Z nuclei, and the corresponding spectroscopic factors.
For odd-Z nuclei, the predicted location of the proton drip line is 
found in excellent agreement with available experimental data, 
and with results calculated with various macroscopic-microscopic
mass models. The even-Z nuclei at the drip line are not
accessible in experiments, and therefore we have compared 
the RHB (NL3+D1S) results with the predictions of various
mass models.

We have also addressed the important issue of ground state
proton radioactivity below $Z=50$. In this region nuclei
beyond the drip line exist only as short lived resonances,
and no examples of ground state proton emitters have been
discovered so far. It is expected that emitted protons could
be observed in the range of transition energies: 
$0.2 - 0.3$ MeV (for Z=30) and $0.35 - 0.5$ MeV (for Z=40).
The present RHB model calculations have shown that a number
of odd-Z nuclei can be expected below $Z=50$, with negative
one-proton separation energy in this energy windows. For
these possible ground state proton emitters we have also
determined the Nilsson orbitals occupied by the odd valence
proton and the corresponding spectroscopic factors.
These results can be used in calculations of partial 
proton decay half-lives, and therefore in comparison
with future experimental data.

\bigskip
\bigskip

\leftline{\bf ACKNOWLEDGMENTS}

This work has been supported in part by the
Bundesministerium f\"ur Bildung und Forschung under
project 06 TM 979, by the Deutsche Forschungsgemeinschaft,
and by the Gesellschaft f\" ur Schwerionenforschung (GSI) Darmstadt.

\newpage

\begin{figure}
\caption{
The proton drip line in the region $31 \leq Z \leq 49$.
On this section of the chart of the nuclides the 
last bound isotopes for each element are indicated. 
Nuclei to the left are predicted to be proton unstable
by the present RHB (NL3+D1S) calculation.}
\label{figA}
\end{figure}

\begin{figure}
\caption{
The proton drip line in the region $35 \leq Z \leq 49$.
The mass number of the first proton unbound nucleus along each 
isotopic chain is plotted as a function of the atomic number.
The results of the present RHB (NL3+D1S) calculation are
compared with the predictions of various mass models.} 
\label{figB}
\end{figure}

\begin{figure}
\caption{
Calculated ground-state quadrupole deformations for the nuclei 
at the proton drip line $31 \leq Z \leq 49$.}
\label{figC}
\end{figure}

\begin{figure}
\caption{
Predictions of the RHB model for the one-proton
separation energies of odd-Z nuclei $31 \leq Z \leq 39$, at and beyond the
drip-line.}
\label{figD}
\end{figure}

\begin{figure}
\caption{
Same as in Fig. \ref{figA}, but for the odd-Z isotopes $41 \leq Z \leq
49$.}
\label{figE}
\end{figure}

\begin{figure}
\caption{
Self-consistent ground-state quadrupole deformations for odd-Z
nuclei $31 \leq Z \leq 47$, at and beyond the proton drip-line.}
\label{figF}
\end{figure}

%%%%%%%%%%%%%%%%%%%%%%%%%%%%%%%%%%%%%%%%%%%%%%%%%%%%%%%%%%%%%
\newpage
\begin{table}
\caption{Candidates for odd-Z ground-state proton emitters 
in the region of nuclei with 
$31\leq Z \leq 49$. The results of RHB calculation for the one-proton
separation energies $S_p$, quadrupole deformations $\protect\beta_2$, and
the deformed single-particle orbitals occupied by the odd valence proton
are compared with predictions of the macroscopic-microscopic mass model. 
All energies are in units of MeV;
the RHB spectroscopic factors are displayed in the sixth column.}
\label{TabA}
\begin{center}
\begin{tabular}{llllllllll}
& N & $S_p$ & $\beta_2$ & $p$-orbital & $u^2$ & $\Omega^{\pi}_p$ \cite
{MNK.97} & $S_p$ \cite{MNK.97} & $\beta_2$ \cite{MN.95}  \\ 
\hline
$^{63}$As &30 & -0.45 & 0.22 & $1/2^-[310]$ & 0.61 & $3/2^-$ & -1.36 & 0.22
&  \\ 
$^{64}$As & 31 & -0.12 & 0.23 & $1/2^-[310]$ & 0.66 & $3/2^-$ & -0.22 & 0.23
  \\ 
$^{68}$Br & 33 & -0.26 &-0.28 & $9/2^+[404]$ & 0.80 & $9/2^+$ & -0.33 &-0.32
\\ 
$^{69}$Br & 56 & -0.10&-0.29 & $9/2^+[404]$ & 0.78 & $9/2^+$ &  0.09 &-0.32
\\ 
$^{72}$Rb & 35 & -0.83 &-0.37 & $7/2^+[413]$ & 0.83 & $7/2^+$ & -0.80 &-0.38
 \\ 
$^{73}$Rb & 36 & -0.31 &-0.34 & $7/2^+[413]$ & 0.83 & $3/2^+$ & -0.31 & 0.37
\\ 
$^{75}$Y & 36 & -0.56 & 0.42 & $5/2^+[422]$ & 0.92 & $5/2^+$ & -1.57 & 0.41
&  \\ 
$^{76}$Y & 37 & -0.03 & 0.41 & $5/2^+[422]$ & 0.84 & $5/2^+$ & -0.57 & 0.41
&  \\ 
$^{81}$Nb & 40 & -0.10& 0.49 & $1/2^+[431]$ & 0.12 & $1/2^+$ & -1.00 & 0.46
&  \\ 
$^{84}$Tc & 41 & -0.55 &-0.21 & $5/2^+[413]$ & 0.90 & $5/2^+$ & -0.76 &-0.22
&  \\ 
$^{85}$Tc & 42 & -0.34 &-0.02 & $9/2^+[404]$ & 0.73 &$ 3/2^+$ & -0.66 & 0.05
&  \\ 
$^{88}$Rh & 43 & -0.87 & 0.04 & $7/2^+[413]$ & 0.51 & $5/2^+$ & -0.70 & 0.05
&  \\ 
$^{89}$Rh & 44 & -0.38 & 0.01 & $9/2^+[404]$ & 0.56 & $5/2^+$ & -0.50 & 0.05
&  \\ 
$^{92}$Ag & 45 & -0.50 & 0.02 & $7/2^+[413]$ & 0.39 & $7/2^+$ & -0.67 & 0.05
\\ 
$^{93}$Ag & 46 & -0.11 & 0.04 & $9/2^+[404]$ & 0.49 & $7/2^+$ & -0.49 & 0.05
\\ 
$^{96}$In &47  & -0.91& 0.04 &$9/2^+[404]$  & 0.19 & $9/2^+$ & -0.38& 0.05
\\
$^{97}$In &48  &  -0.37& 0.02 &$9/2^+[404]$  & 0.21 & $9/2^+$ & -0.34& 0.05

\end{tabular}
\end{center}
\end{table}
%%%%%%%%%%%%%%%%%%%%%%%%%%%%%%%%%%%%%%%%%%%%%%%%%%%%%%%%%%%%%
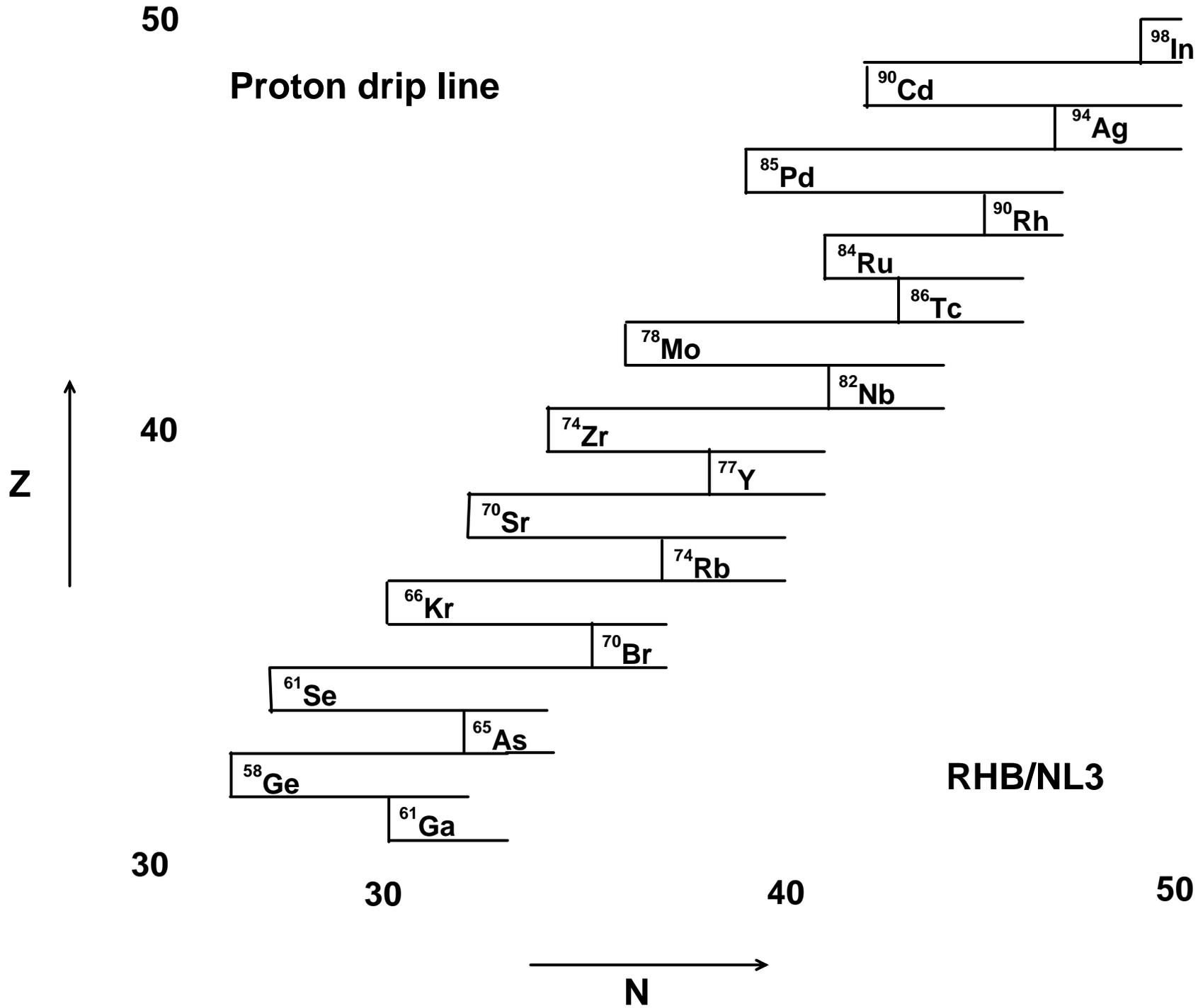
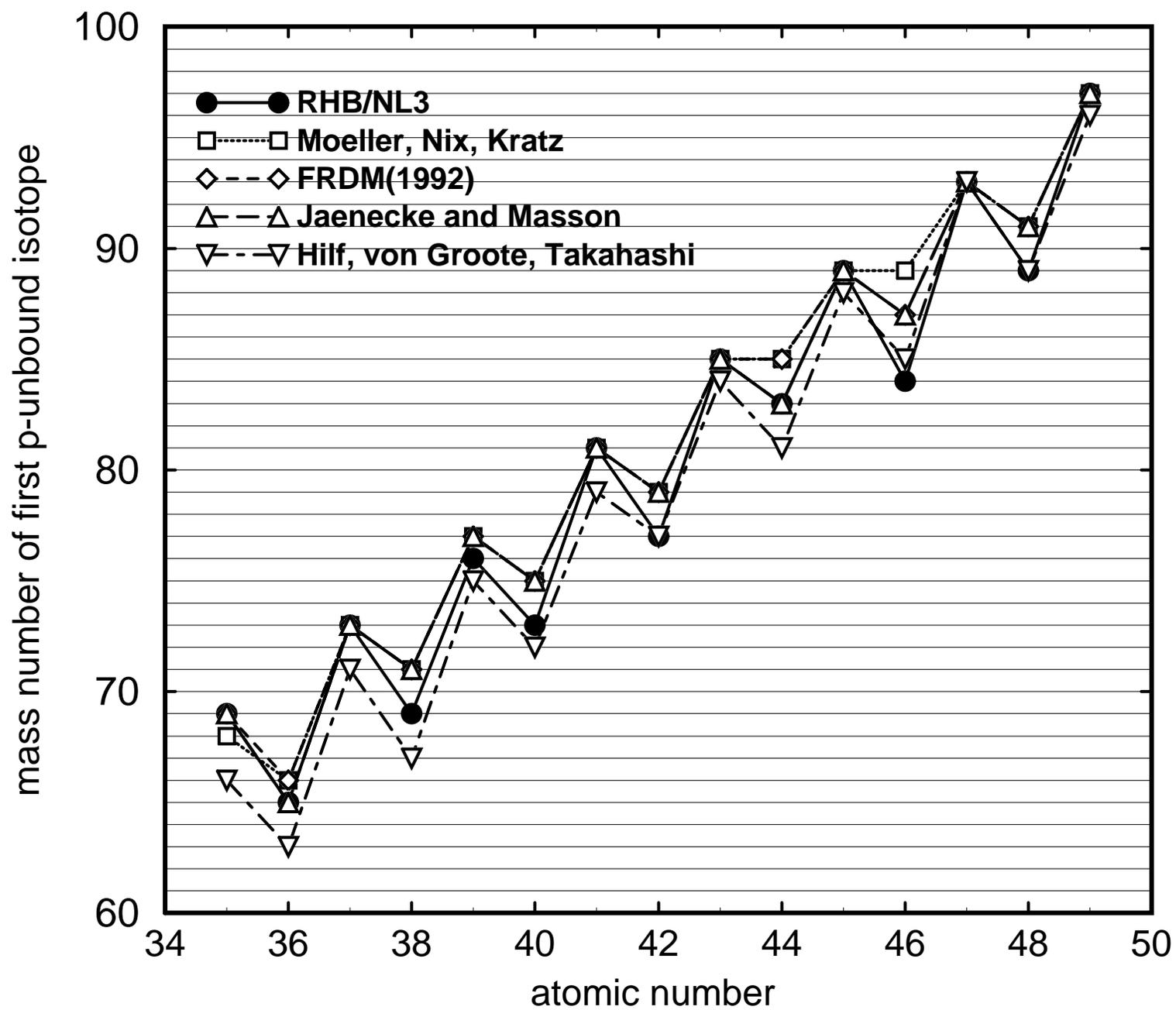
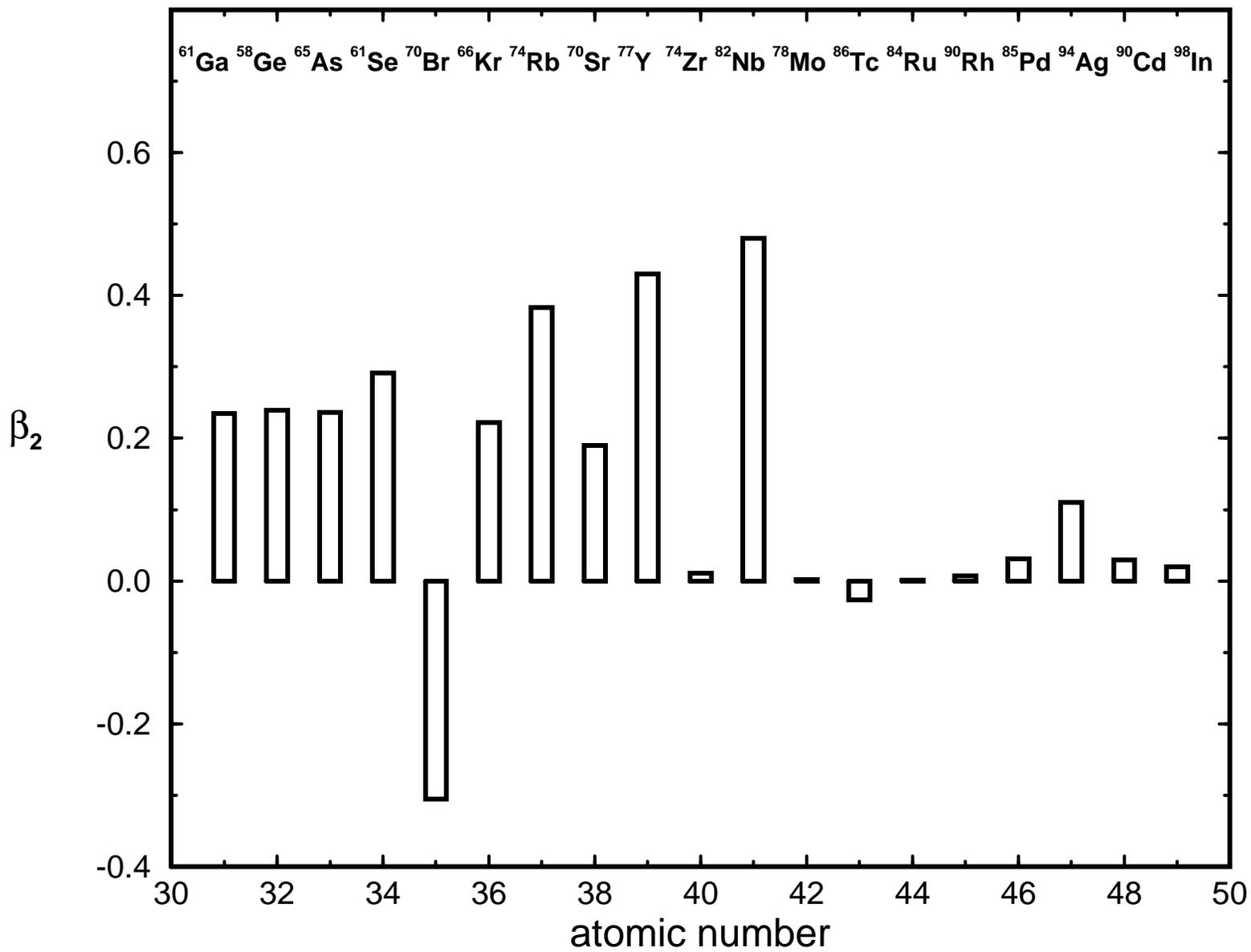
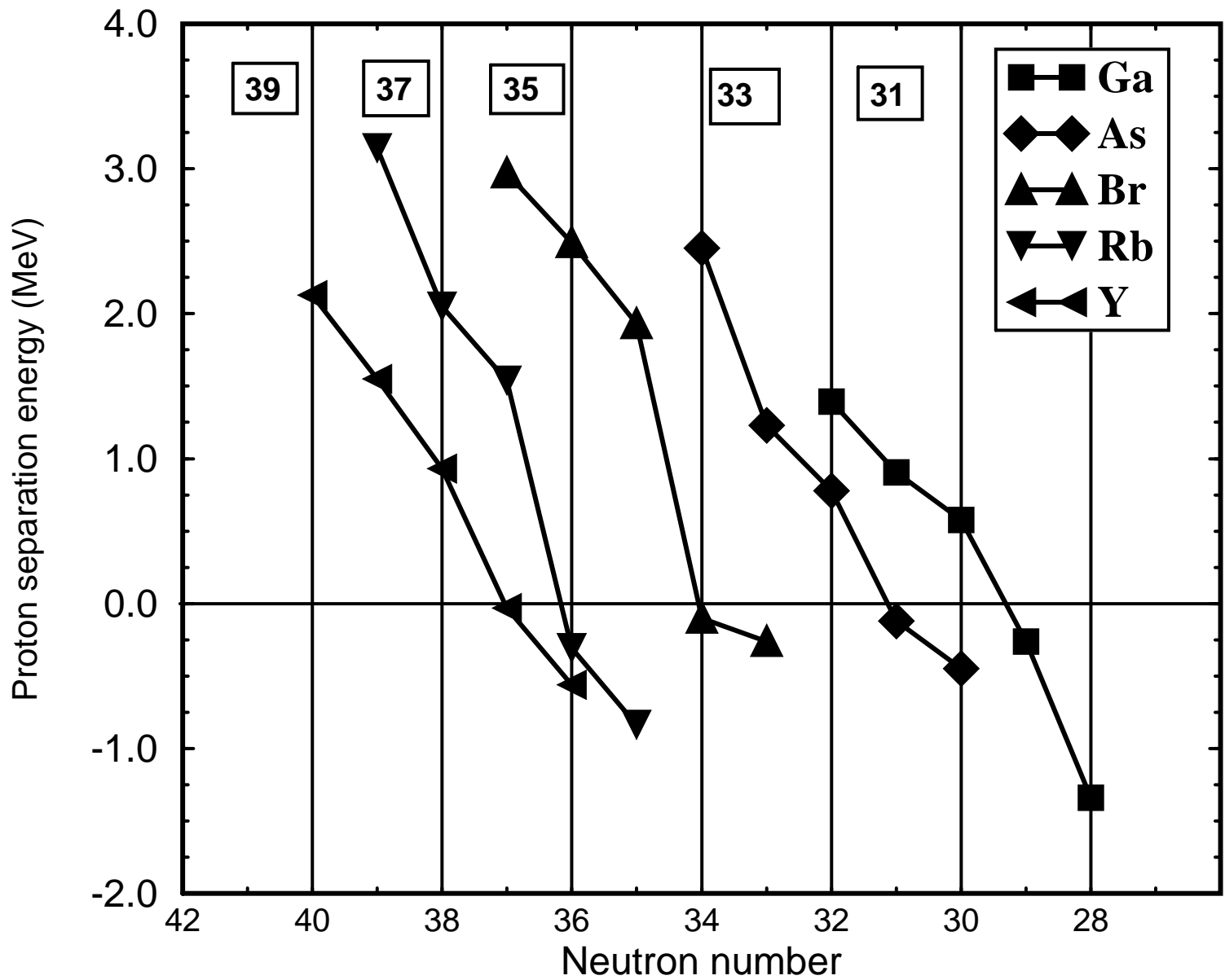
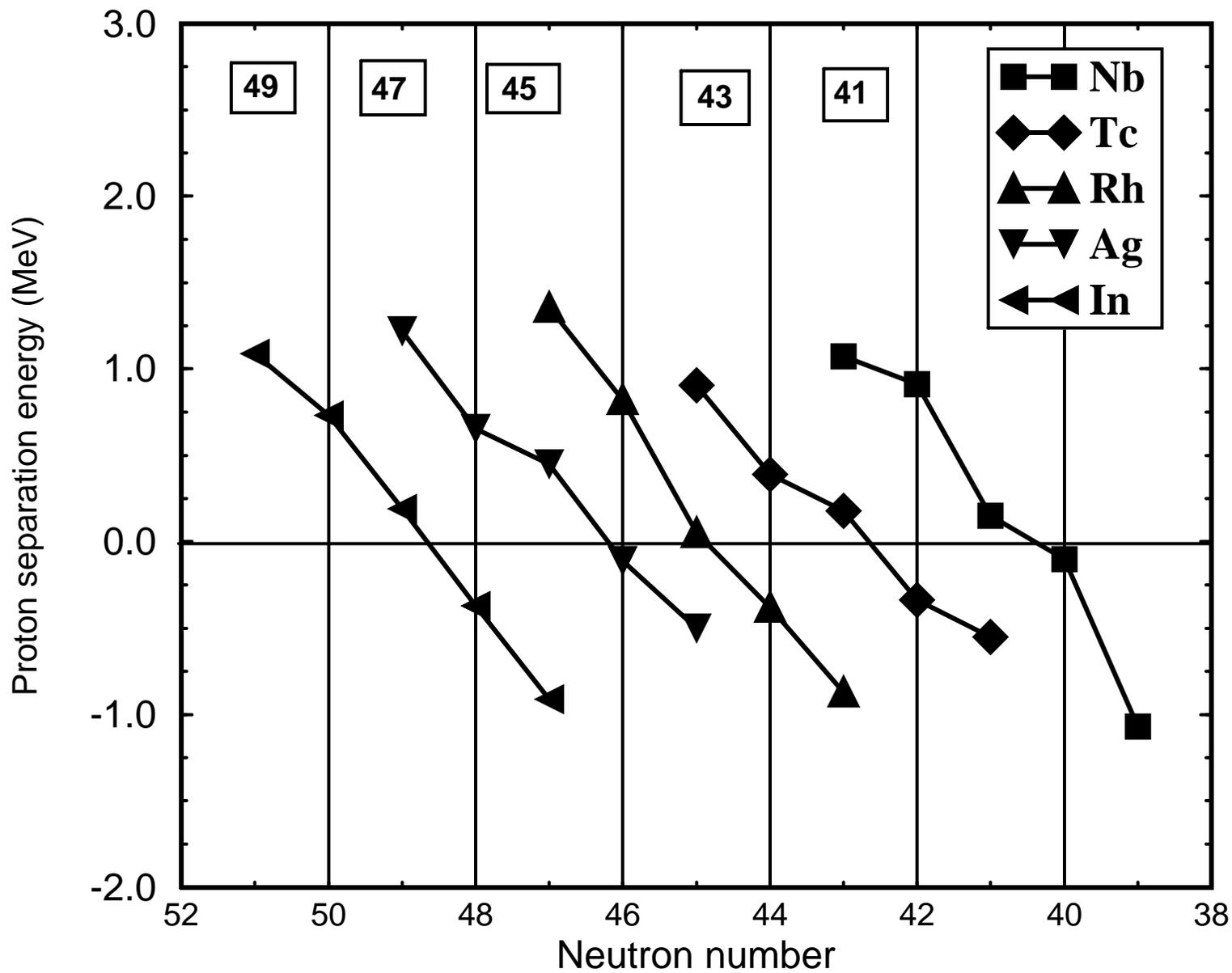
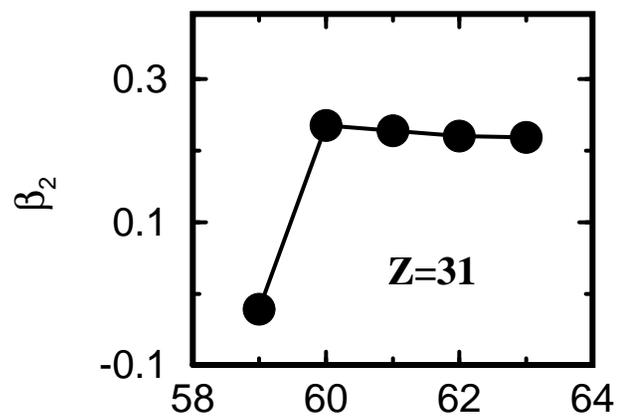
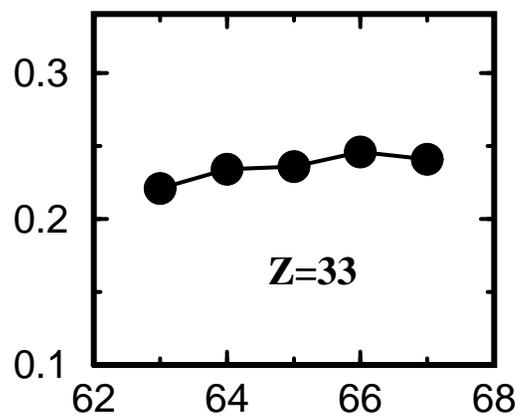
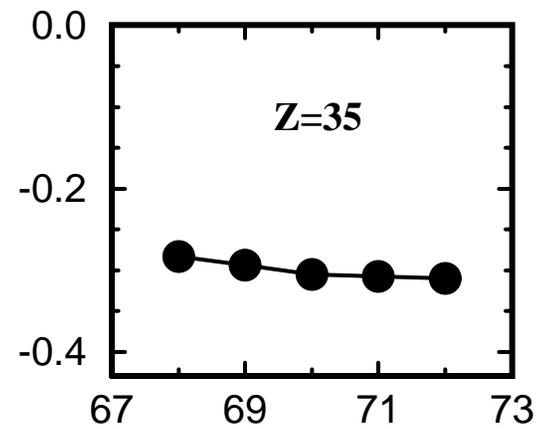
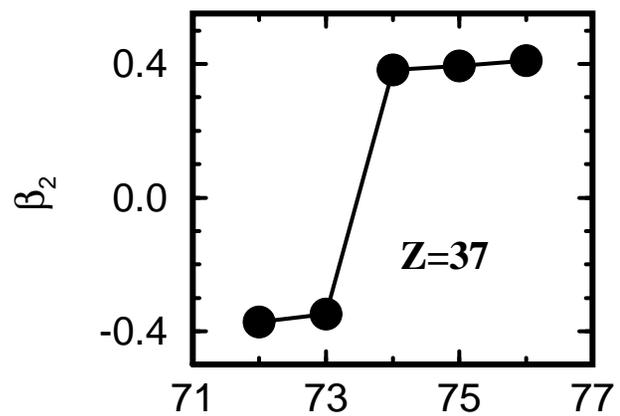
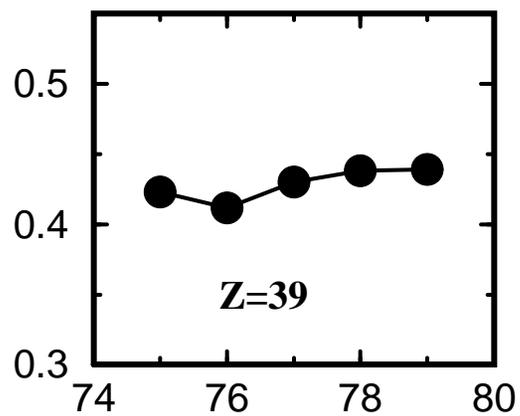
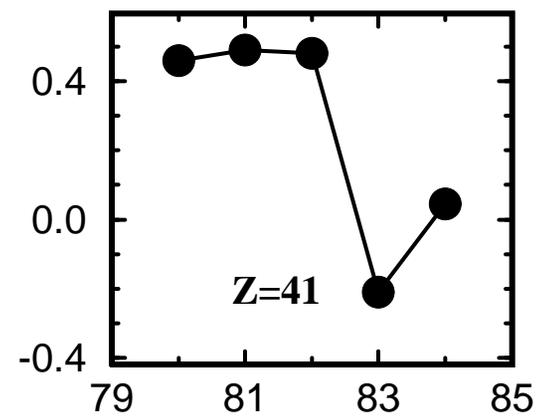
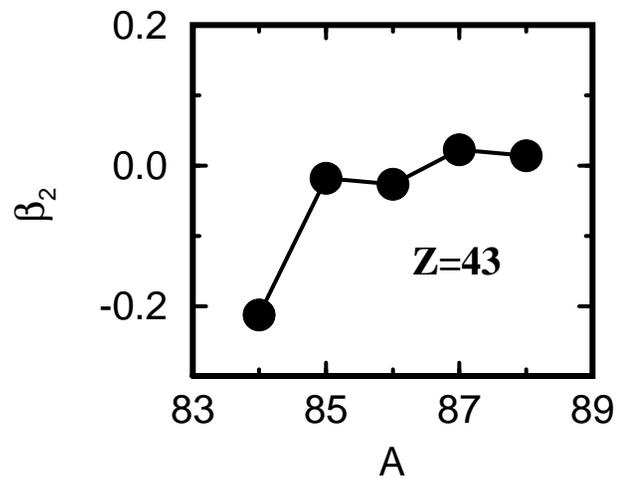
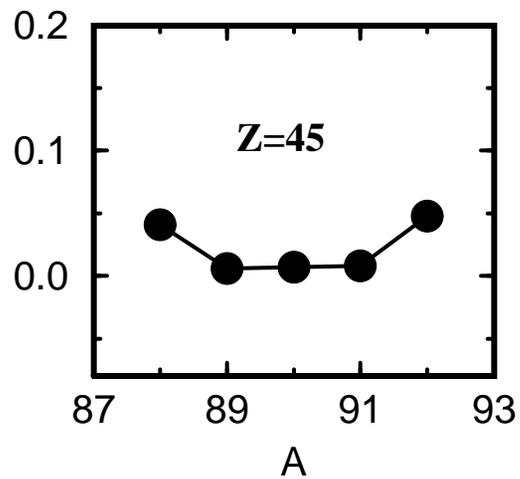
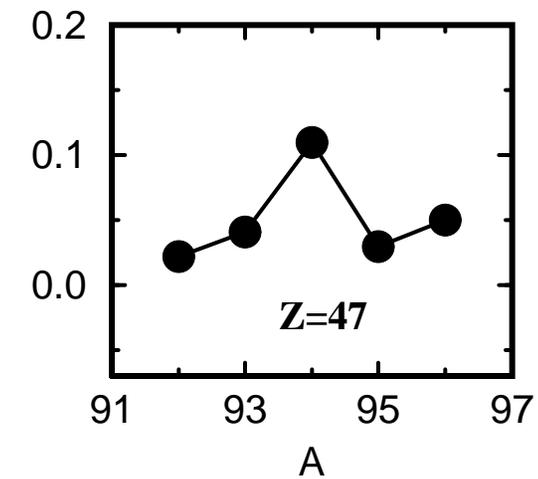
\end{document}